\documentclass[fleqn,usenatbib]{mnras}
\usepackage[T1]{fontenc}
\usepackage{graphicx}
\usepackage{amsmath}
\title[Gas transport and rotation curves in NGC 3198]{Local morphology and asymmetry of galactic rotation curves in a kinetic gas transport framework: NGC 3198}

\author[A. A. Lipovka and A. A. Lipovka]{
Anton A. Lipovka,$^{1}$\thanks{E-mail: anton.lipovka@gmail.com; ORCID: 0000-0003-2770-2304}
and Anna A. Lipovka$^{2}$\\
$^{1}$Physics Research Department, Sonora University, Hermosillo, Sonora, Mexico\\
$^{2}$Colegio Munoz, Hermosillo, Sonora, Mexico
}

\date{Accepted XXX. Received YYY; in original form ZZZ}
\pubyear{2026}

\begin{document}
\label{firstpage}
\pagerange{\pageref{firstpage}--\pageref{lastpage}}

\maketitle

\begin{abstract}
We investigate whether the observed fine structure and asymmetry of non-averaged galactic rotation curves can be reconstructed directly from the observed HI distribution within the framework of a kinetic gas transport description. Using the observed HI density profiles separately for the approaching (north-eastern) side and the receding (south-western) side of the galaxy NGC~3198, we reconstruct the corresponding rotation curves based on the equation previously derived in \citet{Lipovka2022}. It is shown that the reconstructed curves reproduce not only the approximately flat large-scale behaviour of the observed rotation curves, but also their detailed local morphology and asymmetry separately for the north-eastern and south-western sides of the galactic disk. The obtained results indicate that the local structure of galactic rotation curves is closely connected with the local HI distribution and arises naturally as a consequence of kinetic gas transport processes in galactic disks.
\end{abstract}


\begin{keywords}
dark matter --
galaxies: ISM --
galaxies: kinematics and dynamics --
galaxies: spiral --
galaxies: structure --
hydrodynamics
\end{keywords}

\section{Introduction}

The interpretation of galactic rotation curves (RCs) remains one of the central problems of modern astrophysics. Within the standard paradigm, the approximately flat behaviour of RCs at large galactocentric distances is usually interpreted as evidence for the existence of extended dark matter (DM) halos surrounding spiral galaxies. In this context, one of the major problems is the observed strong empirical correlation between the observed distribution of baryonic matter and the shape of galactic RCs, known as "Renzo's rule" \citep{Sancisi1999,Sancisi2004,McGaugh2004}. For example, \citet{McGaugh2004} states that: "The distribution of baryonic mass is completely predictive of the distribution of dark matter, even in dark matter dominated LSB galaxies." In another work, \citet{McGaugh2016} emphasize the discovery of "a correlation between the radial acceleration traced by rotation curves and that predicted by the observed distribution of baryons". The authors note that such a correlation may indicate the existence of previously unaccounted dynamical laws, which is in good agreement with the results obtained in \citet{Lipovka2022}. As one possible solution, \citet{McGaugh2004} suggest giving preference to MOND. However, it should be noted that several observational studies have reported galaxies with unexpectedly small amounts of dark matter or even its apparent absence (for example \citep{vanDokkum2018}). Since MOND cannot operate in some galaxies while being absent in others, questions arise regarding the universality of both the standard dark halo interpretation and MOND-like approaches.

Another important problem concerns the detailed local morphology of observed rotation curves in the outer regions of galactic disks. Although dark matter models and MOND-like approaches successfully reproduce the approximately flat large-scale behaviour of rotation curves in many galaxies, they generally describe only smooth averaged profiles. However, observed rotation curves contain fine structure, local peaks, and asymmetries (differences between the two opposite sides of the galactic gas disk). Unfortunately, these features are often partially suppressed by standard averaging procedures and are usually not taken into account.

Already quite long ago, in HI studies of spiral galactic disks, \citet{Bosma1981} reported a strong empirical correlation between the radial distribution of neutral hydrogen and the observed rotation curves. Within the standard paradigm (under the assumption that the rotation curve is predominantly determined by an additional dynamical component traditionally interpreted as dark matter), explaining this effect is difficult, since DM does not interact electromagnetically with the baryonic component.

Recent weak-lensing observations suggest that approximately flat circular velocities may persist up to scales of several hundred kiloparsecs and possibly even $\sim 1$ Mpc \citep{Mistele2024}. Such behaviour further strengthens the long-standing tension between observed galactic dynamics and conventional finite-size dark matter halo models, while simultaneously reinforcing the empirical coupling between baryonic matter and galactic kinematics.

In the previous work \citet{Lipovka2022}, it was shown that the dynamics of outer galactic HI disks require the inclusion of gas kinetics. In that work, a modified diffusion equation was derived directly from the kinetic equation, and it was shown that it naturally leads to flat rotation curves without the need to introduce dark matter into the model.

The present work develops this approach further and focuses on the local morphology of non-averaged rotation curves. We investigate whether the observed fine structure and asymmetry of rotation curves can be reconstructed directly from the observed HI distribution within the kinetic framework developed in \citet{Lipovka2022}. Thus, the interpretation of rotation curve formation proposed in the present work fundamentally differs from those mentioned above. Within the kinetic approach, the observed correlation between the morphology of the radial HI distribution and the morphology of the rotation curve naturally arises through the kinetic equation, since gas dynamics itself directly contributes to the observed velocity field. In this case, local features of the rotation curves arise from the kinetic properties of the galactic disk gas itself.

Special attention in the present work is devoted to the fine structure of observed rotation curves. Unfortunately, in many studies, RCs from opposite sides of the disk are combined (effectively averaged) in order to obtain a smoother total rotation profile. However, such averaging may partially suppress important information related to the local dynamical structure and asymmetry of the galactic disk.

As a test object, we separately consider the north-eastern and south-western sides of the galaxy NGC 3198. It is shown that the extended dynamical model, taking gas kinetics into account, reproduces not only the global flat behaviour of the rotation curve, but also its detailed local structure and asymmetry.


\section{Standard interpretation and combined model}

As is well known, within the standard DM interpretation, the observed RCs in galactic disks are interpreted within classical mechanics and are assumed to correspond directly to circular orbital motion:

\begin{equation}
\frac{v^2(R)}{R}=\frac{GM(R)}{R^2},
\end{equation}

which is assumed to make it possible to reconstruct the total mass distribution of a galaxy from the observed rotation curve. Within this approach, the discrepancy between the observed approximately flat RCs and the expected Keplerian decline is interpreted as evidence for the existence of an additional invisible mass component identified as DM. It should be emphasized that such an interpretation of RCs within the DM paradigm implicitly assumes that the outer galactic gas behaves as an ensemble of collisionless particles moving along nearly circular orbits and obeying Newtonian mechanics.

As already mentioned above, an additional difficulty for models based on smooth DM halos is related to the observed fine structure and asymmetry of non-averaged rotation curves. Since DM halos are assumed to be smooth and approximately symmetric, local small-scale variations in the baryonic gas density should not produce corresponding local variations in the rotational velocity field, and the rotation curves would be expected to remain symmetric and flat to good accuracy. Reproducing the observed local morphology, branch-specific asymmetry, and small-scale features is therefore not straightforward within conventional smooth-halo models and generally requires additional dynamical assumptions.

In \citet{Lipovka2022}, it was shown that the assumption of collisionless gas is not valid for the outer regions of HI disks ($R > R_{25}$), where kinetic gas transport effects provide the dominant contribution to the dynamics of the galactic disk. Thus, an extended dynamical model of galactic disks must include not only Newtonian mechanics, but also gas kinetics. Such an approach leads to a fundamentally different and more complete physical picture, since the considered system (the galactic disk) becomes dynamically closed. Within this extended framework, local variations of the gas density distribution, through the kinetic equation, lead to corresponding local variations in the rotational velocity profile, whose morphology is in good agreement with observations.

In the aforementioned work, it was also shown that the relation between the radial gas density distribution and the tangential velocity in the outer galactic regions can be naturally explained within the framework of gas kinetics \citep{Lipovka2022}. Since the kinetic equation is rather cumbersome for practical modeling, a more convenient relation connecting the radial gas density distribution with the corresponding rotational velocity was derived there:

\begin{equation}
\frac{D}{n}\frac{1}{R_{0}}\frac{\partial n}{\partial r}=-\sqrt{V_{d\perp}V_{\perp }^{tot}},
\label{eq2}
\end{equation}

where $D$ is the effective diffusion coefficient, $n(R)$ is the radial HI density distribution, $R_0$ is the characteristic radial scale, and $V_{d\perp}$ and $V_\perp^{tot}$ denote the diffusive and total transverse velocity components introduced in \citet{Lipovka2022}. The relation was derived from the stationary kinetic transport equation for a rarefied galactic gas under the assumption of quasi-stationary radial transport.

We emphasize that Eq.~(\ref{eq2}) does not suffer from the restrictions imposed on hydrodynamical equations and therefore can be applied to arbitrary gas densities, which is a necessary condition for describing the dynamics of very rarefied galactic gas disks. A detailed derivation of Eq.~(\ref{eq2}) from the stationary kinetic equation was presented in \citet{Lipovka2022}. In the same work, a reconstruction of the global flat behaviour of RCs was performed based on this equation. In the present work, Eq.~(\ref{eq2}) is used for a more detailed reconstruction of rotation curves from the observed radial HI distributions in order to investigate the small-scale fine structure and asymmetry of RCs and to compare the obtained results with observations.

On the one hand, numerical estimates (see \citet{Lipovka2022}) show that in the regions $R > R_{25} \approx 4'$ of the considered galactic disk, the contribution of baryonic gravity becomes very small compared to the contribution produced by gas kinetics (this is intuitively clear from the fact that at large distances Keplerian curves for galaxies are always much smaller than the observed rotation curves). On the other hand, we are not interested here in describing the relatively narrow transition region near $R \approx R_{25}$, where both Newtonian dynamics and kinetic equations should be included simultaneously. Instead, we focus on sufficiently distant regions of the gas disk. This allows us to simplify the model and consider exclusively the gas kinetic contribution described by Eq.~(\ref{eq2}), neglecting gravitational interaction. We note here that including gravity would only increase the dynamical coupling of the galactic disk (which works in favour of the proposed approach). However, since gravity becomes small at large distances, neglecting it is a reasonable approximation for the present calculations.

Of course, when modelling RCs in the transition region, the full system of equations should be considered (including both Newtonian mechanics and gas kinetics). In the present work, however, the construction of a complete galactic model was not the goal, since this would distract from the main purpose of the study. Therefore, we restrict ourselves to modelling the disk rotation curve beyond $R_{25}$, i.e. in the region where gas motion is predominantly determined by the kinetic equation.


\section{HI profiles}

The observed HI profiles and their Gaussian approximations are presented in Figs.~1 and 2 for the north-eastern and south-western sides of the galactic disk, respectively. The corresponding approximation coefficients are listed in Tables~1 and 2.

The observed radial density distributions were approximated by a sum of Gaussian components,

\begin{equation}
N_{L}(R)=\sum_{k} a_k \exp\left[-b_k(R-c_k)^2\right],
\end{equation}

where the coefficients $a_k$, $b_k$, and $c_k$ were determined by fitting the observational data. The approximation results together with the observed HI profiles are presented in Fig.~1 for the north-eastern side and in Fig.~2 for the south-western side.

As can be seen from Figs.~1 and 2, the observed HI distributions demonstrate noticeable local inhomogeneous structure as well as a clear asymmetry between the north-eastern and south-western sides of the galactic disk. Within our approach, these inhomogeneities lead, through Eq.~(\ref{eq2}), to corresponding details in the fine structure of the reconstructed rotation curves discussed below.

\begin{figure}
\centering
\includegraphics[width=\hsize]{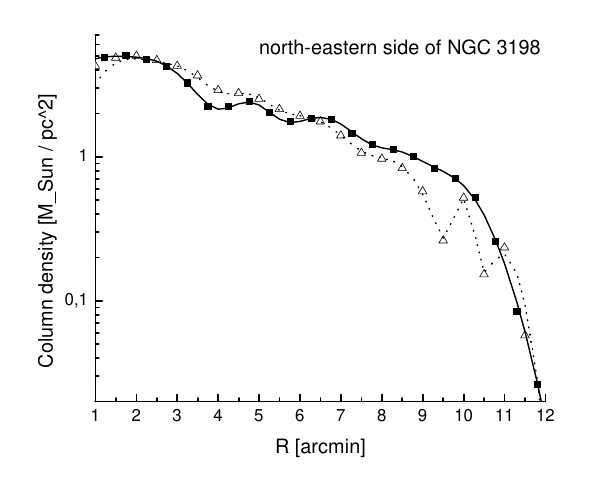}
\caption{Observed radial HI column densities and their Gaussian approximations for the north-eastern side of NGC 3198. Black squares correspond to the observations of \citet{Dove1994}, while open triangles correspond to the observations of \citet{Gentile2013}. The solid line represents the Gaussian approximation of the 1994 observations, while the dashed line corresponds to the approximation of the 2013 data.}
\label{fig1}
\end{figure}

\begin{table}
\footnotesize
\centering
\caption{Gaussian approximation coefficients for the north-eastern HI profile.}
\label{table1}
\centering
\begin{tabular}{c c c c c c c}
\hline\hline
\multicolumn{4}{c}{Data for \citep{Dove1994}} & \multicolumn{3}{c}{Data for \citep{Gentile2013}}\\
\hline
$k$ & $a_k$ & $b_k$ & $c_k$ & $a_k$ & $b_k$ & $c_k$\\
\hline
1 & 2.88 & 0.93 & 0.75 & 3.40 & 1.00 & 1.45\\
2 & 4.48 & 0.42 & 2.35 & 3.95 & 0.59 & 2.97\\
3 & 1.68 & 1.58 & 4.76 & 2.00 & 1.49 & 4.83\\
4 & 1.85 & 0.55 & 6.53 & 0.34 & 2.80 & 5.80\\
5 & 0.53 & 1.51 & 8.36 & 1.57 & 0.90 & 6.43\\
6 & 0.70 & 0.65 & 9.55 & 0.86 & 0.83 & 8.26\\
7 & --- & --- & --- & 0.43 & 10.0 & 10.01\\
8 & --- & --- & --- & 0.21 & 3.00 & 10.98\\
\hline
\end{tabular}
\end{table}

\begin{figure}
\centering
\includegraphics[width=\hsize]{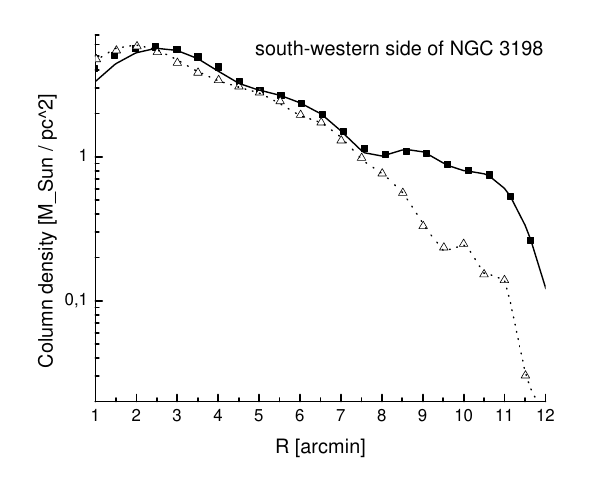}
\caption{Observed radial HI column densities and their Gaussian approximations for the south-western side of NGC 3198. Black squares correspond to the observations of \citet{Dove1994}, while open triangles correspond to the observations of \citet{Gentile2013}. The solid line represents the Gaussian approximation of the 1994 observations, while the dashed line corresponds to the approximation of the 2013 data.}
\label{fig2}
\end{figure}

\begin{table}
\footnotesize
\centering
\caption{Gaussian approximation coefficients for the south-western HI profile.}
\label{table2}
\centering
\begin{tabular}{c c c c c c c}
\hline\hline
\multicolumn{4}{c}{Data for \citep{Dove1994}} & \multicolumn{3}{c}{Data for \citep{Gentile2013}}\\
\hline
$k$ & $a_k$ & $b_k$ & $c_k$ & $a_k$ & $b_k$ & $c_k$\\
\hline
1 & 5.66 & 0.21 & 2.59 & 5.80 & 0.32 & 1.85\\
2 & 1.21 & 0.99 & 5.37 & 2.43 & 0.34 & 4.71\\
3 & 1.40 & 0.84 & 6.58 & 0.20 & 2.70 & 5.33\\
4 & 1.06 & 0.74 & 8.70 & 0.85 & 1.50 & 6.65\\
5 & 0.67 & 0.85 & 10.59 & 0.65 & 0.87 & 8.04\\
6 & --- & --- & --- & 0.04 & 7.70 & 9.10\\
7 & --- & --- & --- & 0.22 & 2.70 & 10.01\\
8 & --- & --- & --- & 0.13 & 7.90 & 10.90\\
9 & --- & --- & --- & 0.023 & 7.80 & 11.50\\
10 & --- & --- & --- & 0.015 & 5.50 & 12.20\\
\hline
\end{tabular}
\end{table}

In conclusion of this section, we emphasize once again that the observed radial HI distributions exhibit both (i) fine structure and (ii) asymmetry.


\section{Reconstructed rotation curves}

Using the approximated HI density distributions and Eq.~(\ref{eq2}) derived in \citet{Lipovka2022}, we calculated the corresponding rotation curves separately for the north-eastern and south-western sides of NGC~3198. The reconstructed rotation curves for the north-eastern and south-western sides are presented in Figs.~3 and 4.

\begin{figure}
\centering
\includegraphics[width=\hsize]{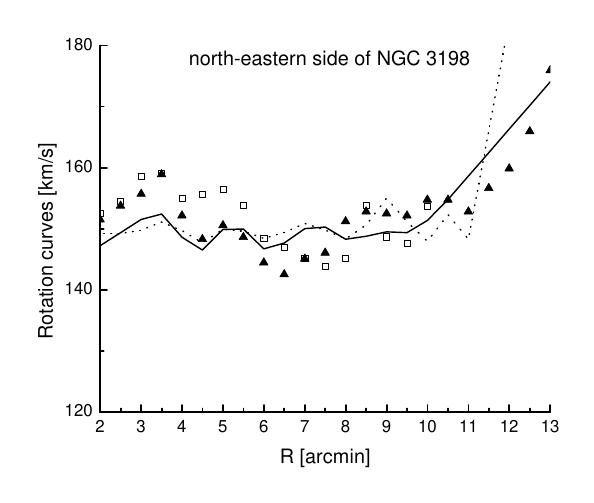}
\caption{Observed and reconstructed rotation curves for the north-eastern side of NGC 3198. Open squares correspond to the observations of \citet{Begeman1987}, while black triangles correspond to the observations of \citet{Gentile2013}. The solid line represents the rotation curve reconstructed using Eq.~(\ref{eq2}) from the HI profiles presented in \citet{Dove1994}, while the dotted line represents the rotation curve reconstructed from the HI profiles measured by \citet{Gentile2013}.}
\label{fig3}
\end{figure}

\begin{figure}
\centering
\includegraphics[width=\hsize]{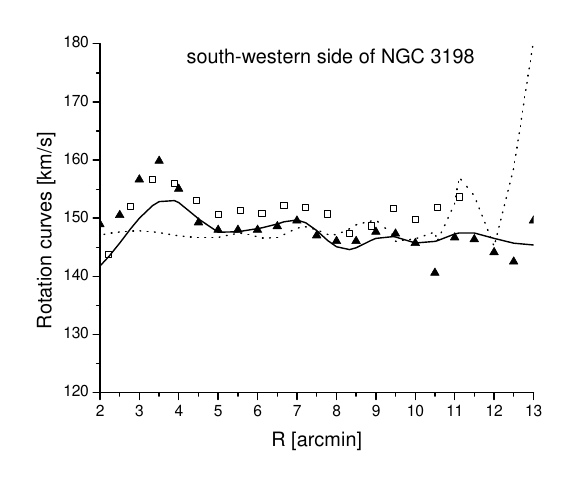}
\caption{Observed and reconstructed rotation curves for the south-western side of NGC 3198. Open squares correspond to the observations of \citet{Begeman1987}, while black triangles correspond to the observations of \citet{Gentile2013}. The solid line represents the rotation curve reconstructed using Eq.~(\ref{eq2}) from the HI profiles presented in \citet{Dove1994}, while the dotted line represents the rotation curve reconstructed from the HI profiles measured by \citet{Gentile2013}.}
\label{fig4}
\end{figure}

As can be seen from Figs.~3 and 4, substantial differences exist between the north-eastern and south-western rotation curves. All local extrema of the RCs observed in the north-eastern and south-western sides are well reproduced by the reconstruction performed within the framework that includes gas kinetics. It should be emphasized that features appearing as local maxima in the north-eastern RC correspond to minima or suppressed structures in the south-western RC, reflecting the asymmetry already initially present in the HI distributions (see, for example, the differences in the RC behaviour at distances $R = 5.2'$ ; $R = 8.4'$ ; $R = 9.2'$ ; $R = 10.3'$ ; $R = 11'$ ; $R = 12'$), which argues in favour of the developed approach. Thus, the north-eastern and south-western structures of the RCs possess different morphologies that are linked to the gas density profiles through Eq.~(\ref{eq2}). Moreover, the north-eastern and south-western structures of the RCs also exhibit different global trends. In particular, the north-eastern side of the RC shows an ascending trend, whereas the south-western side exhibits a descending trend.

The reconstruction obtained here indicates a systematic connection between local gradients of the HI distribution and local extrema of the rotation curves. The reconstructed curves reproduce not only the approximately flat global behaviour of the outer rotation curves obtained in \citet{Lipovka2022}, but also the observed local morphology specific to each side of the RC. We emphasize that the agreement is obtained separately for the north-eastern and south-western sides without introducing DM or additional free parameters required for postulating the configuration of a dark matter halo. This is achieved despite the substantially different local morphology exhibited by the two sides of the galactic disk.


\section{Discussion}


The results obtained in the present work show that the local structure of galactic rotation curves contains important physical information that may be partially suppressed by standard averaging procedures when combining approaching/receding RCs.

As shown previously in \citet{Lipovka2022}, the description of galactic disk dynamics should include both Newtonian gravitational dynamics and the kinetic properties of galactic gas. However, the estimates presented in that work indicate that the contribution of gas kinetics becomes dominant in the outer regions of galactic disks (at distances larger than $R_{25}$), where HI gas dominates.

It should be noted that different observational studies provide somewhat different values of $R_{25}$ for NGC 3198 - from approximately $3.23'$ \citep{Schmidt2016} to $4.25'$ \citep{Kendall2011}. This reflects the fact that $R_{25}$ is not a strictly defined physical boundary, but rather an observationally determined characteristic scale separating the bright inner optical disk, populated predominantly by stars, from the outer HI-dominated region.

Within the framework considered here, this distance naturally corresponds to a gradual transition zone between the inner dynamically orbital regime (where Newtonian mechanics dominates) and the outer regime, where gas dynamics is predominantly governed by the kinetic equation. Clearly, in this transition region the full system of equations (Newtonian mechanics together with the kinetic equation) should be solved. However, in the present work we are interested exclusively in the outer part of the galactic disk, where the contribution of the kinetic equation dominates (see \citet{Lipovka2022}). For this reason, for modelling rotation curves in the regions $R > R_{25} \approx 4'$, it may be a reasonable approximation to use only Eq.~(\ref{eq2}) without including gravity (which would only further increase the coupling of the system). Quantitative estimates supporting this approximation were presented in \citet{Lipovka2022}.

Minimal number of Gaussian components required for the reconstruction is determined by the number of local features present in the observed HI distributions. The reconstruction also indicates the existence of a connection between local gradients of the gas density distribution and local extrema of the rotation curves. Within this approach, the reconstructed rotation curves reproduce not only the approximately flat global behaviour in the outer galactic regions, but also the observed local morphology and asymmetry between the north-eastern and south-western sides.

Within the standard dark matter paradigm, rotation curves are usually modelled using smooth and approximately symmetric halo distributions. In such approaches, reproducing the observed local morphology and branch-specific asymmetry requires additional assumptions or auxiliary dynamical mechanisms. In contrast, within the kinetic-gas dynamical framework, when gas kinetics is consistently taken into account, the observed velocity field becomes closely connected to the local HI distribution. This naturally explains the existence of different fine structures and asymmetries in the approaching and receding sides of the RC prior to their summation, which leads to averaging of the fine structure.

An additional interesting aspect of the kinetic interpretation concerns the recent observational detection of approximately flat rotation curves extending to distances approaching $\sim 1$~Mpc from galactic centres \citep{Mistele2024}. Such scales significantly exceed the optical radii of galaxies and extend far into the outer HI-dominated regions. Within the standard dark matter paradigm, maintaining dynamically stable flat RCs on such enormous spatial scales may require extremely extended dark matter halos with specific density profiles.

Within the kinetic-gas framework considered here, however, such behaviour appears naturally. As shown in \citet{Lipovka2022}, beyond the optical disk ($R \ge R_{25}$), where the HI component becomes dominant, the contribution of gas kinetics exceeds the contribution from baryonic gravitational dynamics. In this regime, the rotation curve is governed primarily by Eq.~(\ref{eq2}), which directly relates the velocity field to the gradient of the radial HI density distribution.

Importantly, both the HI density distribution and its radial derivative entering Eq.~(\ref{eq2}) are smooth continuous functions. Consequently, the kinetic description naturally produces smooth non-Keplerian velocity profiles over very large spatial scales without requiring additional assumptions about extended dark matter halos. In this sense, the existence of approximately flat rotation curves extending to scales approaching $\sim 1$~Mpc may be qualitatively consistent with the kinetic transport interpretation of galactic dynamics. 


\subsection{Limitations of the present reconstruction}


Several limitations of the present reconstruction should be noted.

First, the observed HI profiles and rotation curves used in this work were reconstructed from previously published observational figures, since the original tabulated datasets and corresponding uncertainties are no longer publicly available. Consequently, the present analysis is primarily focused on the morphological correspondence between the observed HI distributions and the branch-specific structure of the reconstructed rotation curves rather than on high-precision quantitative fitting.

Second, the present reconstruction is restricted to the outer HI-dominated regions of the galactic disk ($R > R_{25}$), where the contribution of gas kinetics is expected to dominate according to the estimates presented in \citet{Lipovka2022}. The transition region near $R \approx R_{25}$, where both Newtonian gravitational dynamics and kinetic gas transport should simultaneously contribute, was not modelled self-consistently in the present work.

Third, the model considered here does not include additional physical processes such as magnetic fields, turbulence, stellar feedback, or non-stationary gas flows. The aim of the present study was instead to investigate whether the observed local morphology and asymmetry of non-averaged rotation curves may already emerge naturally within the kinetic transport framework.

Despite these limitations, the reconstructed rotation curves reproduce not only the global behaviour of the observed outer RCs, but also a substantial part of their branch-specific local morphology and asymmetry.

\section{Conclusions}

In the present work, we tested a reconstruction scheme for the detailed morphology of non-averaged galactic rotation curves within the kinetic framework previously proposed in \citet{Lipovka2022}. Using the galaxy NGC 3198 as an example, we have shown that the observed morphology of rotation curves in the outer galactic regions may be naturally explained within a model consistently taking gas kinetics into account.

The observed HI distributions for the north-eastern and south-western sides of the galactic disk were independently approximated and used for the reconstruction of the corresponding rotation curves by means of Eq.~(\ref{eq2}), previously derived from the kinetic equation describing the dynamics of rarefied galactic gas. In this framework, the observed correlation between HI morphology and the local morphology of the rotation curves naturally arises within the kinetic approach. In particular, local extrema observed in the rotation curves systematically correspond to local features (extreme values of the gradient) in the HI distribution. Within the kinetic-gas dynamical framework, the reconstructed rotation curves reproduce not only the approximately flat global behaviour in the outer galactic regions, but also the observed local morphology, the specific branch structure, the asymmetry between the approaching and receding branches, and the global trends (deviations of the rotation curves from flat linear approximations on large scales).

Thus, on the one hand, the numerical estimates presented in \citet{Lipovka2022} indicate the importance of gas kinetics in dynamical processes at large distances from galactic centres, which points to the necessity of including gas kinetics in extended dynamical models of galactic disks. On the other hand, when RCs are reconstructed directly from the observed HI profiles, the calculated RC values agree with the observed ones. Namely, the rotation curves and the radial HI distributions exhibit correlated fine structure, which within conventional smooth-halo DM models usually requires additional assumptions or auxiliary dynamical mechanisms. Moreover, the reconstructed rotation curves exhibit noticeable differences between the approaching and receding sides of the galactic disk, morphologically similar to the observed differences of the rotation curves themselves.

It should be noted that the approaching and receding sides of the RC exhibit different trends. Both for the observed and reconstructed rotation curves of NGC 3198, the approaching side shows an ascending trend, whereas the receding side exhibits a descending trend.

The observed correspondence between asymmetric HI morphology and asymmetric branch-specific RC morphology suggests that part of the observed non-axisymmetric velocity structure is dynamically linked to gas transport processes. Considering that kinetic gas transport provides a substantial contribution to the observed dynamics of outer galactic disks, non-averaged rotation curves contain important physical information about the kinetics of galactic gas disks and may serve as a sensitive observational test for improving models of galactic dynamics. Moreover, the good agreement obtained without a dark matter component suggests that any additional dark matter contribution to the observed velocity field must be constrained.


\section*{Data Availability}

The data underlying this article are available from the sources cited in the text.


\begin{thebibliography}{}

\bibitem[Begeman(1987)]{Begeman1987} Begeman, K. 1987, PhD Thesis, University of Groningen

\bibitem[Bosma(1981)]{Bosma1981} Bosma, A. 1981, AJ, 86, 1825

\bibitem[Dove \& Shull(1994)]{Dove1994} Dove, J. B., \& Shull, J. M. 1994, ApJ, 423, 196

\bibitem[Gentile et al.(2013)]{Gentile2013} Gentile, G., Jozsa, G. I. G., Serra, P., et al. 2013, A\&A, 554, A125

\bibitem[Kendall et al.(2011)]{Kendall2011}
Kendall, S., Kennicutt, R. C., \& Clarke, C. 2011, MNRAS, 414, 538

\bibitem[Lipovka(2022)]{Lipovka2022} Lipovka, A. 2022, Int. J. Mod. Phys. A, 37, 2250171

\bibitem[McGaugh(2004)]{McGaugh2004} McGaugh, S. S. 2004, ApJ, 609, 652

\bibitem[McGaugh et al.(2016)]{McGaugh2016} McGaugh, S. S., Lelli, F., \& Schombert, J. M. 2016, Phys. Rev. Lett., 117, 201101

\bibitem[Mistele et al.(2024)]{Mistele2024} Mistele, T., McGaugh, S. S., Lelli, F., Schombert, J., \& Li, P. 2024, ApJL, 969, L3, doi:10.3847/2041-8213/ad54b0 

\bibitem[Sancisi(1999)]{Sancisi1999} Sancisi, R. 1999, Astrophys. Space Sci., 269, 59

\bibitem[Sancisi et al.(2004)]{Sancisi2004} Sancisi, R., Fraternali, F., Oosterloo, T., \& van der Hulst, T. 2004, in IAU Symp. 220, Dark Matter in Galaxies, ed. S. Ryder et al., 233

\bibitem[Schmidt et al.(2016)]{Schmidt2016} Schmidt, T. M., Bigiel, F., Klessen, R. S., \& de Blok, W. J. G. 2016, MNRAS, 457, 2642

\bibitem[van Dokkum et al.(2018)]{vanDokkum2018} van Dokkum, P., Danieli, S., Cohen, Y., et al. 2018, Nature, 555, 629

\end{thebibliography}
\end{document}